\documentclass{aastex}

\begin{document}

\title{POWER DENSITY SPECTRA OF GAMMA-RAY BURST LIGHT CURVES: 
       IMPLICATIONS ON THEORY AND OBSERVATION}

\author{Heon-Young Chang and Insu Yi}

\affil{Korea Institute For Advanced Study \\
207-43 Cheongryangri-dong, Dongdaemun-gu, Seoul, 130-012, Korea}
\email{hyc@kias.re.kr, iyi@kias.re.kr}

\begin{abstract}

We investigate statistical properties
of GRB light curves by comparing the reported characteristics in the PDSs 
of the observed GRBs with those that we model, and discuss implications on 
interpretations of the PDS analysis results. 
Results of PDS analysis of observed GRBs suggest that the averaged PDS 
of GRBs follows a power law over about two decades of frequency with 
the power law index, $-5/3$, and the distribution of individual power follows 
an exponential distribution. Though an attempt to identify the most sensitive
physical parameter has been made on the basis of the internal shock model,
we demonstrate that conclusions of this kind of approach should be derived
with due care. We show that the reported slope and the distribution
can be reproduced by adjusting the sampling interval in the time domain for a given
decaying timescale of individual
pulse in a specific form of GRB light curves. In particular,  given that
the temporal feature is modeled by a two-sided exponential function, the power law 
behavior with the index of $-5/3$ and the exponential distribution of the observed PDS
is recovered at the 64 ms trigger time scale when the decaying timescale of individual 
pulses is $\sim 1$ second, provided that the pulse sharply rises. 
Another way of using the PDS analysis is an application of the same method to 
individual long bursts in order to examine a possible evolution of the decaying
timescale in a single burst.

\end{abstract}

\keywords{gamma rays:bursts -- methods:numerical}

\section{INTRODUCTION}

Since 16 gamma-ray bursts (GRBs) were discovered in the late sixties by {\it Vela}
satellites \citep{kle73}, several satellites have been dedicated to observe
the bursts and numerous theories were suggested to explain their nature and origin
\citep{nar92,mes93,woo93,rees94,usov94,yi97,black98,pac98,fry99,mac99,port99}.
Unfortunately, however, the origin of GRBs has remained unsettled for more than 
three decades. Observations
of the afterglow of GRBs enable us to establish the facts that GRBs are cosmological
\citep{mao92,meegan92,pi92,met97}
and the emission of the afterglow is due to the electron  synchrotron radiation from 
a decelerating relativistic blast wave \citep{pac93,sari95,wax97a,wax97b,wij97},
which suggests indirect  hints of  the emission mechanism of GRBs. 
From the observations of several GRB afterglows the evidence of beamed GRBs has 
accumulated \citep{sari99, hal99, rhoads99}, and there are works on models for 
the geometry of GRBs (e.g., Mao and Yi 1994).
Durations of GRBs range from about 30 ms to about 1000 s, and show a bimodality 
in the logarithmic distribution \citep{fish95}. 
On the contrary, study of the afterglows (e.g., Piran 1999 and references therein) 
deals with the emission on much longer timescales
(e.g., months, or even up to years) than GRB emission timescales. 
This is both good and bad for the subject. It is good because details of 
the complicated initial conditions are largely irrelevant to the calculation.
It is bad because the study of the afterglow reveals very limited information
on the central engine of the GRBs. 

To examine proposed GRB theories one has to consider the following points : 
observed isotropy and inhomogeneity in space, apparent flux
distribution, temporal and spectral features observed in bursts. 
Among others, the study of burst morphologies is a difficult task because of diversity,
apparently no clear correlation with other observational parameters, relatively
undeveloped methods for the study of temporal structures of GRBs. Nonetheless,
there are several attempts to quantify pulse shapes of GRBs and interpret results
in terms of physics \citep{fen96,nor96,zand96,kob97,bel98,dai98,fen99,pan99}. 

The Fourier transform technique is widely used to study hydrodynamical turbulence 
and to search  for the underlying process in the system  as well as periodical 
phenomena \citep{brace65}. 
In most of the GRB models, an individual burst is a random realizaiton of a single
stochastic process. Features of such a process can be probed with statistical 
methods applied to a sufficiently large ensemble. Provided that the GRBs are
generated by the same origin, one may employ the simplest statistical quantity, 
i.e., the average. A possible way to subtract statistical fluctuations from the 
underlying characteristics is to take the average of PDSs over samples. 
Then the fluctuations affecting each individual PDS tend to cancel out each other 
and one can see the underlying features.
\citet{bel98} applied the Fourier transform technique 
to the analysis of 214  light curves of long GRBs ($T_{90} > 20$ sec). 
They found that, even though individual PDSs were very diverse the 
averaged PDS was in accord with a power law of index $-5/3$ over 2 orders of magnitude
of a  frequency range, and that fluctuations in the power were distributed according to 
the 
exponential distribution. They also noted that the  value of the slope was the same
as the Kolmogorov spectrum  of velocity fluctuation in a turbulent fluid. They 
concluded that the GRB emission was generated in a relativistic and fully developed
turbulent outflow, resulting from the coalescence of two neutron stars or a neutron
star and a black hole.  However, they implicitly assume that the
selected bursts ($ T_{90} > 20~ {\rm sec}$) are long enough compared with 
the temporal resolution  (e.g., $64~ $ms) and that the rise and decay time scales
 have no effects on the resulting slope of the PDSs.  

 \citet{pan99} analyzed the temporal behavior of GRBs in the framework of a relativistic 
internal shock model, using the power density spectrum. They set up 
their internal shock model, and attempted to identify the most sensitive model
 parameters to the PDS and to explore the 
efficiency of conversion of kinetic energy of shells to radiation. They suggested that
the wind must be modulated such that collisions at large radii release more energy
than those at small radii in order to reproduce the consistent PDSs with the observation.

We address the following questions : Is the sampling interval of 64 ms in the time domain
really short enough to obtain bias-free conclusions in the PDS analysis? 
Or can one see the decaying  timescale if the slope and the sampling timescale are given,
provided that there is a relation among those parameters? Answers to these questions
may well have implications on interpretations of PDS analysis results, such as, those 
in \citet{bel98}, and an evolution of GRB light curve during the GRB emissions.
One may take the PDSs of GRB light curves in separate energy channels 
instead of bolometric light curves as we do in this {\it Letter}. 
It is well-known that the pulses in 
a GRB are more narrow in a higher energy band. One therefore expects that 
the averaged PDS has different slopes in different channels for a given 
sampling interval due to different timescales in different energy bands.
Evidence for a broad luminosity function is found when looking at the 
isotropic luminosities of the bursts with measured redshifts.  The issue of 
whether the dim bursts are intrinsically weak remains unsettled yet.
The difference of bright and dim bursts in temporal behaviors may be an 
important fact in this respect. One may study a correlation between the burst 
brightness and the PDS slope.
In this {\it Letter}, we demonstrate that the Fourier transform technique can 
be used in investigations of the behavior of the 'central engine'. 
We construct a simple model for GRBs, for simplicity, considering a two-sided 
exponential function (see Norris et al. 1996).

\section{PDS OF ARTIFICIAL GRB LIGHT CURVES}

Light curves of GRBs show the diverse temporal profiles. Besides differences in
different bursts, pulse shapes exhibit a broad range in a form of individual 
pulse, in the rise and decay time scales, in a variability. 
Burst asymmetry on short time scales results from the tendency for most 
($\sim$ 90~\%) pulses to rise more quickly 
than they decay, the majority having rise-to-decay time scale ratios of 
$0.3 - 0.5$, independent of energy. Nonetheless, it is worth noting that 
not all of the bursts show FRED shape (Fast Rise, Exponential Decay). 
Some of GRBs show symmetric pulse shapes,
or even reversed behaviors, that is, slow rise and fast decay. 
The dominant trend of spectral softening 
seen in most pulses arises partially from faster  onsets at higher energy and 
slower decays at lower energies, although in addition, pre-cursors appear in 
the higher energy band in some cases.

Even though shapes for all pulses within a single 
burst show variations from pulse to pulse 
\citep{nor96}, we describe GRB light curves as a sum of two-sided exponential functions
given by:
\begin{eqnarray}
f(t)= \sum_m f_m(t),
\end{eqnarray}
where
\begin{eqnarray}
f_m(t)&=&\Lambda_m\exp(a_m(t-t_m)),~~~~~ t<t_m \\
     &&\Lambda_m\exp(-b_m(t-t_m)),~~~~~ t>t_m,\nonumber
\end{eqnarray}
$\Lambda_m$ being the height of peaks, $t_m$ being the time of the pulse's maximum
intensity, $a_m^{-1}$ and $b_m^{-1}$ being the rise and decay timescales, respectively.
Then, the Fourier transform of the function is obtained analytically. 
Since the Fourier transform is a linear operator, the Fourier 
transform of $f(t)$ is a sum of the Fourier transforms of $f_m(t)$, $F_m(\omega)$, 
which reads
\begin{eqnarray}
F_m(\omega)&=&\Lambda_m\int^{t_m}_{0} \exp(a_m(t-t_m)+i\omega t)dt \\
         &&+ \Lambda_m\int^{T}_{t_m} \exp(-b_m(t-t_m)+i\omega t)dt,\nonumber
\end{eqnarray}
where $i=\sqrt{-1}$, $T$ is  the observational duration, or the duration of the burst, 
and $\omega$ is the angular frequency. Having done the integration we have
the final resulting PDS is given by 
\begin{eqnarray}
P(\omega)&=&\sum_m \sum_n F_m(\omega) F_n^{*}(\omega)\\
&=&\sum_m \sum_n \Bigl[\frac{\Lambda_m\Lambda_n \exp(-(a_m t_m+a_n t_n))}{(a_m a_n+\omega^2)^2 + \omega^2(a_m-a_n)^2}\\
&&~~~\{(a_m a_n + \omega^2)g_1(\omega)+\omega(a_m-a_n) g_2(\omega)\}\nonumber\\
&&+\frac{\Lambda_m\Lambda_n \exp(b_m t_m+b_n t_n)}{(b_m b_n+\omega^2)^2 + \omega^2(b_m-b_n)^2}\nonumber\\
&&~~~\{(b_m b_n + \omega^2)g_3(\omega)+\omega(b_m-b_n)g_4(\omega)\}\nonumber\\
&&+\frac{2\Lambda_m\Lambda_n \exp(-a_m t_m+b_n t_n)}{(\omega^2-a_m b_n)^2 + \omega^2(a_m+b_n)^2}\nonumber\\
&&~~~\{( \omega^2- a_m b_n)g_5(\omega)+\omega(a_m+b_n)g_6(\omega)\} \Bigr],\nonumber
\end{eqnarray}
where $g_k(\omega)$'s are complicated  $\cos$ and $\sin$ terms which cause fluctuations on
the PDS. 

In practice, however, we sample 
GRB light curves every pre-determined time interval, e.g., 64 ms. The time interval 
defines the Nyquist frequency, which limits the region we see the information in the frequency
domain  \citep{brace65}. Therefore, unless the sampling interval is short enough 
compared with the typical decaying timescale, the
resulting PDS cannot reveal generic features of the PDS of the original function
in the time domain. For instance, consider the PDS of the bi-exponential function.
Basically, the PDS of the exponential function is given by
\begin{eqnarray}
P(\omega)&\approx&\frac{1}{a^2+\omega^2},
\end{eqnarray}
where $a^{-1}$ is the typical decaying timescale.  However, if an observer takes
insufficiently frequent samples, that is, the Nyquist frequency
is not sufficiently large, then one may see the transition region of the PDS from
the flat part to the power law part with the slope of $-2$. 
The power is dominated 
upto $\omega \sim a$, where $\omega = 2 \pi \nu$. The PDS appears flat until 
$\nu \sim a /2 \pi$, and falls with the slope of -2 as one may expect. 
A summation of the bi-exponential function with randomly distributed decaying timescales 
smears out the transition region of the PDS. 
The slope one may end up with is not 
determined analytically in that the region and it is determined randomly 
around a typical
value of the decaying timescale. We empirically obtain a conclusion that
a sampling interval which yields unbiased slope of the PDS should be 
smaller than the decaying time scale by at least a couple of orders.

\section{RESULTS}

We have generated 100 artificial light curves of  GRBs in the frequency domain 
for random rising and decaying constants $a_m, b_m$ and the waiting time between peaks
$\Delta t_m$. The number of peaks is about 20 in our artificial data. 
The duration of the bursts is fixed to 20 seconds.
We consider noise-free signal. Unless there is 
a systematically biased noise in data, the noise can be regarded as 'white'. And 
effects of this kind of noise should be irrelevant for our conclusions, since
 errors due to such a white noise will be averaged out after all. 
Once generating the Fourier transform of each light curve 
we take a square of its modulus to obtain an individual PDS, then we average PDSs. 
Before taking the Fourier transform of 
light curves we scale them such that the height of their highest peak has unity in the 
artificial GRB light curves.
This has been done to compare  our results with those of \citet{bel98}.
We find that our conclusions are insensitive to 
adopted statistics of $a_m$, $b_m$, and $\Delta t_m$.

In Figure 1, we show that the average of 100 PDSs of our model. What is shown in 
Fig. 1 is essentially the same PDS, but in different parts of the PDS 
in the frequency domain. As the
sampling interval becomes shorter, the Nyquist frequency becomes larger. Subsequently, 
the maximum frequency in the  plots becomes larger for more frequent sampling.
Different parts of the PDS appears to follow a slightly different slope.
For comparison, dotted lines with a slope of $-2$ and dashed lines with $-5/3$ are 
shown. 
For given rising and decaying timescales, the slope of the average PDS is subject to
the sampling interval in the time domain. The $-5/3$ slope is no longer universal for 
the PDS of such an artificial light curve. 
Instead the  observed slope of the averaged PDS should be 
considered as a function of the rising and decaying timescales, and the sampling 
interval. Even for the PDS analysis of long bursts ($T_{90} > 20$ sec), which is longer
than the shortest triggering time scale ($64$ ms) in three orders of magnitude, the
currently available triggering timescale may not be short enough to be free from 
a possible bias.
In Figure 2, the distribution of individual powers is shown. The dashed line is 
the theoretical exponential distribution. The distribution of individual powers
almost exactly follows the exponential distribution.

\section{DISCUSSION}

Based on the fact that one may recover the slope of the PDS of the artificial light 
curves and the distribution of powers by adjusting the rising and decaying 
timescales and the sampling interval we conclude that the observed slope is 
ambiguous. Further more, unless one resolves the issue as to whether the 
currently available time interval is short enough, in comparison with the rising and 
decaying timescales, efforts to identify a controlling parameter on the behavior 
of the PDS should be carried out with due care. As we have demonstrated, 
a conclusion from such kind of analysis is not unique.

Can one determine the decaying time scale for a given sampling interval and 
observed slope? The answer to this question is certainly yes, only provided that
the light curve is properly modeled. As seen in the observational data (e.g.,
Norris et al. 1996), the temporal features of the GRBs are diverse. One way to 
practically use the Fourier transform method is to divide light curves according to
a similarity. In other words, one may select bi-exponential looking light curves
and bi-gaussian looking light curves, and apply a different model function to
each group separately to obtain the timescales.

Provided that one implements a sophisticated algorithm to accommodate the diversity of
the light curves with further efforts, this method could be used for  more important
problems such as the evolution of the GRB emission in a single burst and the
classification of the origin of short and long bursts. The two classes may have 
intrinsically different flare time scales which could be identified in an analysis
similar to the present one. For a
long burst, there is an argument that the GRB emission mechanism is not necessarily
unique even in the same burst (e.g., Yi and Blackman 1997). 
One would like to apply this method to long bursts individually, 
and see whether there is a signature of a possible evolution in the GRB emission mechanism.
One may also explore the emission timescale of long bursts and short bursts to account
for two different origins of them, for instance, as suggested by \citet{yi98}.

\acknowledgments

We thank C. Kim and K. Kwak for useful discussions.
IY is supported in part by the KRF grant No. 1998-001-D00365.

\clearpage

\figcaption[fig1.ps]{The average of 100 PDSs of the model. For comparison, 
dotted lines corresponding to a slope, $-2$, 
and dashed lines to $-5/3$ are shown. 
From left to right, the sampling intervals are 1 sec, 64 ms, $10^{-3}$ sec.
In these plots, $a_m$ and $b_m$ are around 
4.5 ${\rm sec^{-1}}$ and 1.5 ${\rm sec^{-1}}$, respectively. 
\label{fig1}}

\figcaption[fig2.ps]{The distribution of individual powers is shown. The dashed line is 
the theoretical exponential distribution. The corresponding sampling interval
is 64 ms, the rise and decay timescales are same as in Fig. 1. \label{fig2}}

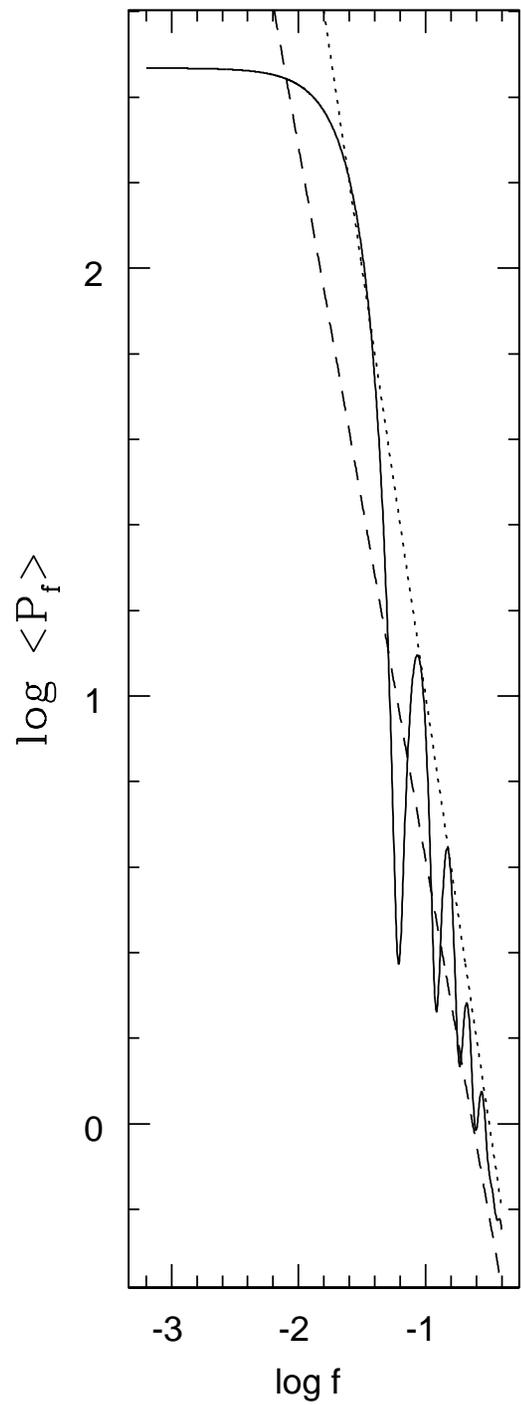
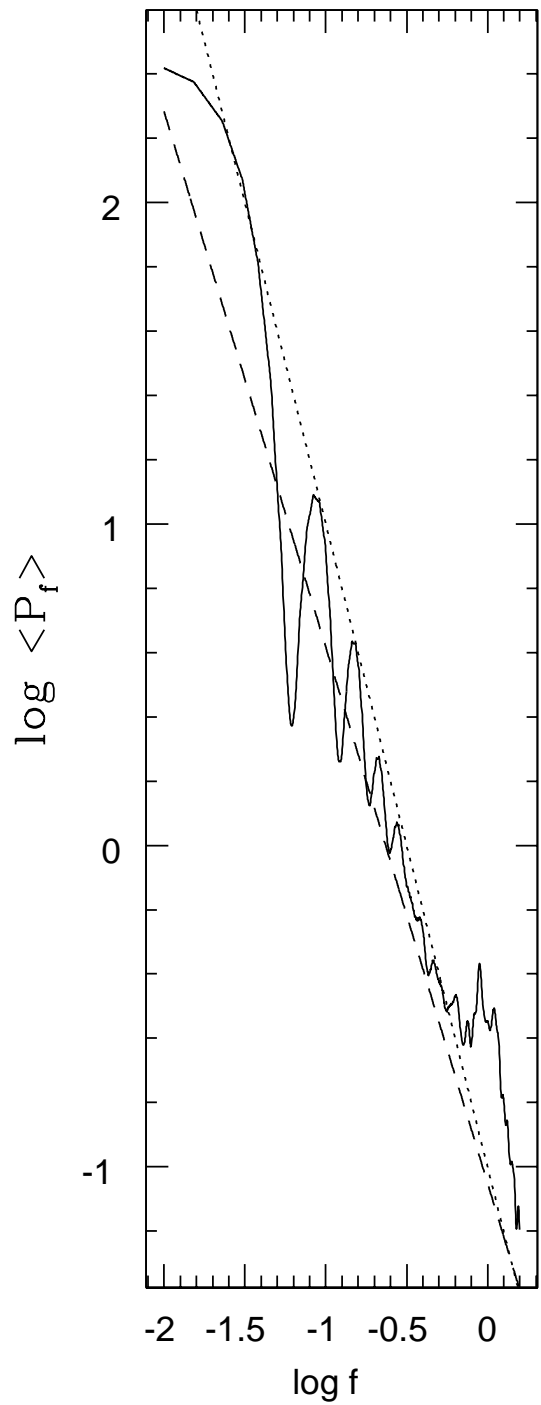
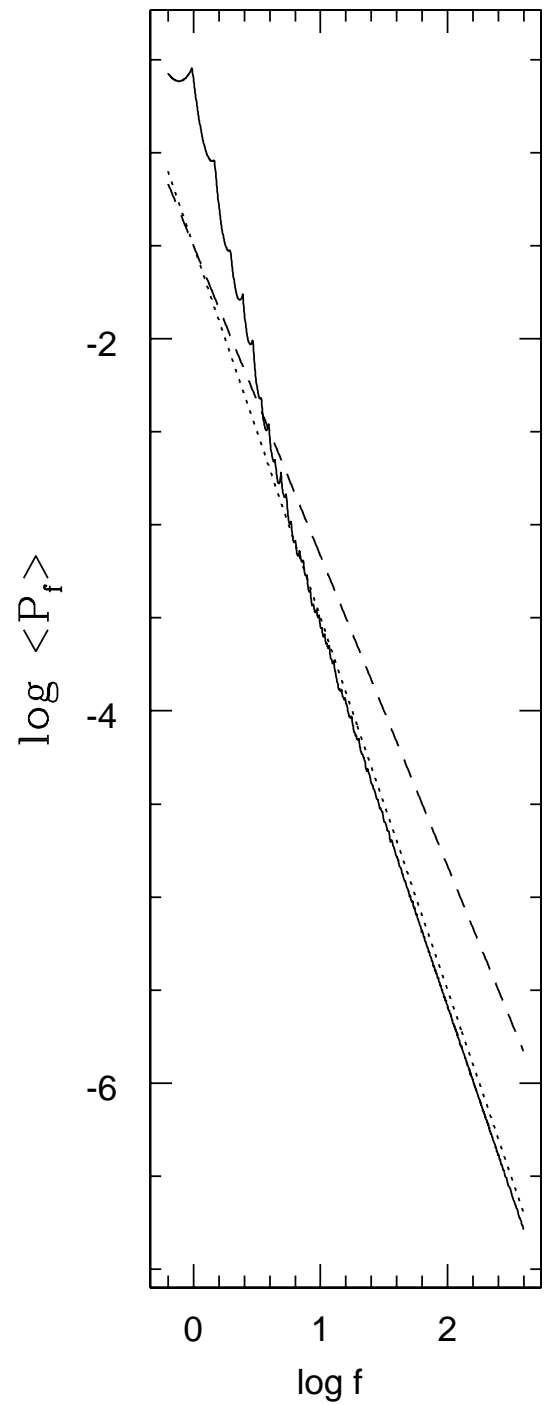
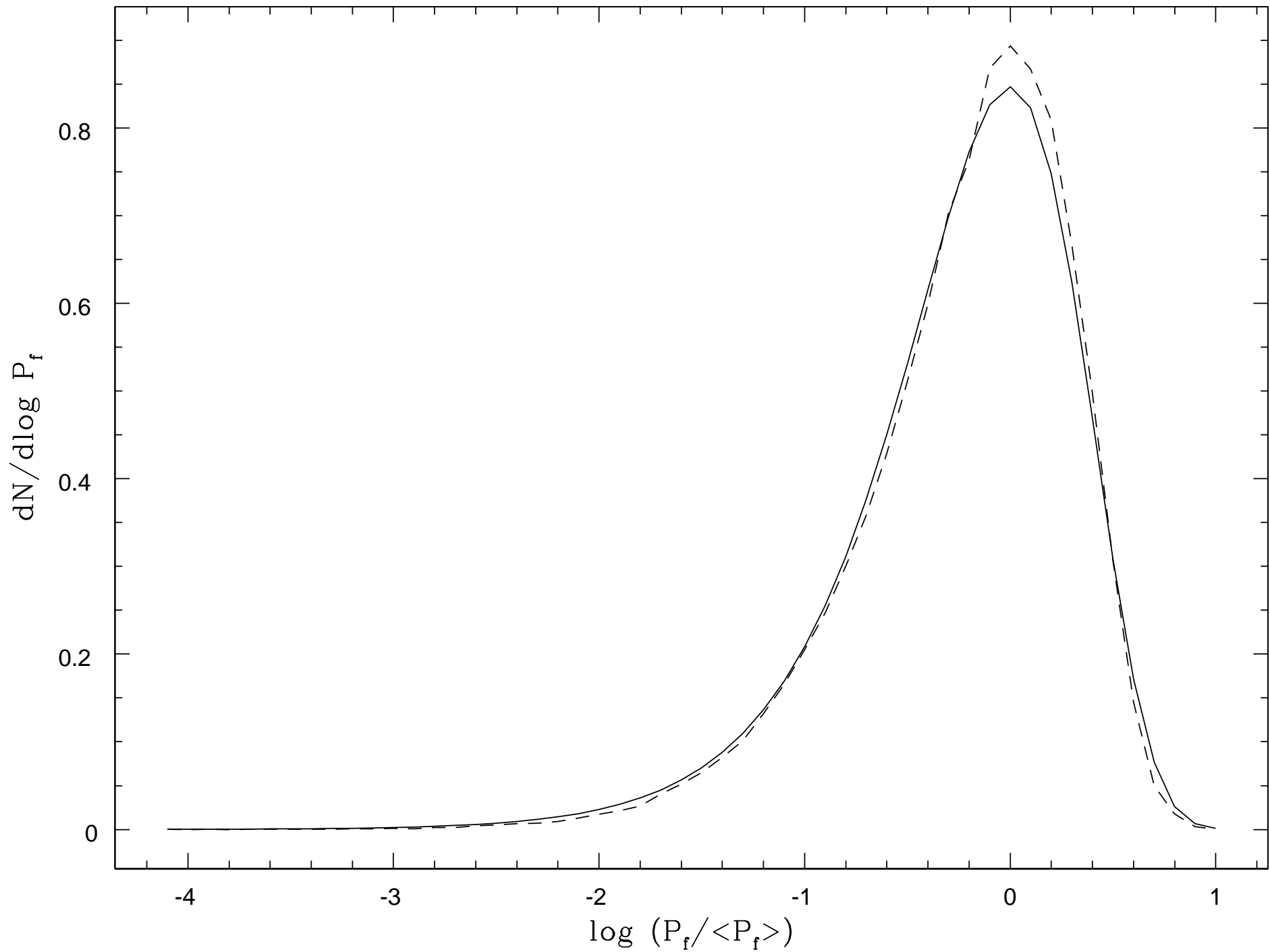
\end{document}